\documentclass[11pt]{article}
\usepackage{graphicx,amsmath,amsfonts,bm,url,color,fullpage}
\usepackage{algorithmic,algorithm}

\newcommand{\R}{\mathbb{R}}

\newcommand{\POs}[1]{{\cal P}_{\Omega_{#1}}}

\newcommand{\diag}{\operatorname{diag}}

\newcommand{\eq}[1]{(\ref{eq:#1})}

\title{Online Identification and Tracking of Subspaces from Highly Incomplete Information}

\author{Laura Balzano$^{\sharp}$, Robert Nowak$^{\sharp}$ and Benjamin Recht$^{\dagger}$\\
  \vspace{-.1cm}\\
  $\sharp$ Department of Electrical and Computer Engineering.\\
University of Wisconsin-Madison\\
  \vspace{-.3cm}\\
  $\dagger$ Department of Computer Sciences. University of Wisconsin-Madison }

\date{October 1, 2010}

\begin{document}

\maketitle

\begin{abstract}
This work presents GROUSE (Grassmannian Rank-One Update Subspace Estimation), an efficient online algorithm for tracking subspaces from highly incomplete observations.  GROUSE requires only basic linear algebraic manipulations at each iteration, and each subspace update can be performed in linear time in the dimension of the subspace.   The algorithm is derived by analyzing incremental gradient descent on the Grassmannian manifold of subspaces.  With a slight modification,  GROUSE can also be used as an online incremental algorithm for the matrix completion problem of imputing missing entries of a low-rank matrix. GROUSE performs exceptionally well in practice both in tracking subspaces and as an online algorithm for matrix completion.
\end{abstract}

\section{Introduction}
The evolution of high-dimensional dynamical systems can often be well 
summarized or approximated in low-dimensional subspaces.  Certain patterns of computer network traffic including origin-destination flows can be well represented by a subspace model~\cite{lakhina04}.  Environmental monitoring of soil and crop conditions~\cite{gupchup07}, water contamination~\cite{papadimitriou05}, and seismological activity~\cite{Wagner96} have all been demonstrated to be efficiently summarized by very low-dimensional subspace representations.    

The subspace methods employed in the aforementioned references are based upon first collecting full-dimensional data from the systems and then using approximation techniques to identify an accurate low-dimensional representation.
However, it is often difficult or even infeasible to acquire and process full-dimensional measurements.  For example, collecting network traffic measurements at a very large number of points and fine time-scales is impractical.  An alternative is to randomly subsample the full-dimensional data.  If the complete full-dimensional data is well-approximated by a lower dimensional subspace, and hence is in some sense redundant, then it is conceivable that the subsampled data may provide sufficient information for the recovery of that subspace.  This is the central intuition of our proposed on-line algorithm for identifying and tracking low-dimensional subspaces from highly incomplete (i.e., undersampled) data.  Such an algorithm could enable rapid detection of traffic spikes or intrusions in computer networks~\cite{ lakhina04} or could provide large efficiency gains in managing energy consumption in a large office building~\cite{Gershenfeld10}.

In this paper, we present GROUSE (Grassmannian Rank-One Update Subspace Estimation), a  subspace identification and tracking algorithm that builds high quality estimates from very sparsely sampled vectors. GROUSE implements an incremental gradient procedure with computational complexity linear in dimensions of the problem, and is scalable to very high-dimensional applications.  An additional feature of GROUSE is that it can be immediately adapted to an `online' version of the matrix completion problem, where one aims to recover a low-rank matrix from a small, streaming random subsets of its entries.  GROUSE is not only remarkably efficient for online matrix completion, but additionally enables incremental updates as columns are added or entries are incremented over time.  These features are particularly attractive for maintaining databases of user preferences and collaborative filtering. 


\section{Problem Set-up}\label{sec:setup}
We aim to track an $d$-dimensional subspace of $\R^n$ that evolves over time, denoted by $S[t]$. 
At every time $t$, we observe a vector $v_t\in S[t]$ at locations $\Omega_t \subset\{1,\ldots n\}$.  Let $\POs{t}$ denote the $|\Omega_t|\times n$ matrix that selects the coordinate axes of $\R^n$ indexed by $\Omega_t$. That is, we observe
$\POs{t} v_t$, $t=1,2,\dots$.   We will measure the error of our subspace using the natural cost function: the squared Euclidean distance from our current subspace estimate $S_{\mathrm{est}}[t]$ to the observed vector $v_t$ \emph{only on the coordinates revealed in the set in $\Omega_t$}:
\begin{equation*}
	F(S_{\mathrm{est}}[t],t) := \operatorname{dist}(\POs{t}(S_{\mathrm{est}}[t]),\POs{t}(v_t))^2
\end{equation*} 
We can compute $F(S;t)$ explicitly for any subspace $S$.  Let $U$ be any matrix whose columns span $S$.  Let $U_{\Omega_t}$ denote the submatrix of $U$ consisting of the rows indexed by $\Omega_t$.  For a vector $v 
\in \R^n$, let $v_{\Omega_t}= \POs{t}(v_t)$ denote a vector in $\R^{|\Omega_t|}$ whose entries are indexed by $\Omega_t$.  Then we have
\begin{equation}\label{eq:f-def}
	F(S;t) = \min_{a} \|U_{\Omega_t} a - v_{\Omega_t}\|^2
\end{equation}
We will use this definition in Section~\ref{derivation} to derive our algorithm.    
If the matrix $U_{\Omega_t}^T U_{\Omega_t}$ has full rank,  then we must have that $w = (U_{\Omega_t}^T U_{\Omega_t})^{-1} U_{\Omega_t}^T v_{\Omega_t}$ achieves the minimum in~\eq{f-def}.  Thus, 
 \begin{equation*}
 	F(S;t) = v_{\Omega_t}^T(I - U_{\Omega_t} (U_{\Omega_t}^T U_{\Omega_t})^{-1} U_{\Omega_t}^T)  v_{\Omega_t}\,.
 \end{equation*}

In the special case where the subspace is time-invariant, that
is $S[t]=S_0$ for some fixed subspace $S_0$, then it is natural to consider the average cost function 
\begin{equation}\label{eq:global-opt}
	\bar F(S) := \sum_{t=1}^T \operatorname{dist}(\POs{t}(S),\POs{t}(v_t))^2 \ .
\end{equation} 
The average cost function will allow us to estimate the steady-state behavior of our  algorithm.  Indeed, in the static case, our algorithm will be guaranteed to converge to a stationary point of $\bar F(S)$.

\subsection{Relation to Matrix Completion}\label{sec:mc}

In the scenario where the subspace does not evolve over time and we only observe vectors on a finite time horizon, then the cost function~\eq{global-opt} is identical to the matrix completion optimization problem studied in~\cite{Keshavan10a,Dai10}.  To see the equivalence, let $\Omega = \{(k,t)~:~k \in \Omega_t\, 1\leq t\leq T\}$, and let $V=[v_1,\ldots,v_T]$.  Then 
\begin{equation*}
\begin{split}
	\bar F(S) &=  \sum_{t=1}^T \min_{a} \|U_{\Omega_t} a - v_{\Omega_t}\|^2\\
	 &= \min_{A\in \R^{d\times T}}\sum_{(i,j)\in\Omega} (UA - V)_{ij}^2
\end{split}
\end{equation*}
That is, the global optimization problem can be written as
$	\min_{U,A} \sum_{(i,j)\in\Omega} (UA - V)_{ij}^2$,
which is precisely the starting point for the algorithms and analyses in~\cite{Keshavan10a,Dai10}.  The authors in~\cite{Keshavan10a} use a gradient descent algorithm to jointly minimize both $U$ and $A$ while~\cite{Dai10} minimizes this cost function by first solving for $A$ and then taking a gradient step with respect to $U$.  In the present work, we consider optimizing this cost function one column at a time.  We show that by using our online algorithm, where each measurement $v_t$ corresponds to a random column of the matrix $V$,  we achieve state-of-the-art performance on matrix completion problems.

\section{Stochastic Gradient Descent on the Grassmannian}
\label{derivation}

The set of all subspaces of $\R^n$ of dimension $d$ is denoted $\mathfrak{G}(n,d)$ and is called the~\emph{Grassmannian}.  The Grassmannian is a compact Riemannian manifold, and its geodesics can be explicitly computed~\cite{Edelman98}.   An element $S\in\mathfrak{G}(n,d)$ can be represented by any $n\times d$ matrix $U$ whose columns form an orthonormal basis for $S$.
Our algorithm derives from an application of incremental gradient descent on the Grassmannian manifold.  We first compute a gradient of the cost function $F$, and then follow this gradient along a short geodesic curve in the Grassmannian.

We follow the program developed in~\cite{Edelman98}.  To compute the gradient of $F$ on the Grassmannian manifold, we first need to compute the partial derivatives of $F$ with respect to the components of $U$.  For a generic subspace, the matrix $U_{\Omega_t}^TU_{\Omega_t}$ has full rank provided that $|\Omega_t|>d$, and hence the cost function~\eq{f-def} is differentiable almost everywhere.  Let $\Delta_{\Omega_t}$ be the $n\times n$ diagonal matrix which has $1$ in the $j^{th}$ diagonal entry if $j\in\Omega_t$ and has $0$ otherwise.  We can rewrite 
 \begin{equation*}
 	F(S;t) = \min_a \|\Delta_{\Omega_t}(U a -v_t)\|^2
 \end{equation*}
from which it follows that the derivative of $F$ with respect to the elements of $U$ is
\begin{align}
\frac{dF}{dU} 
		\nonumber & = -2(\Delta_{\Omega_t} (v_t -  U w)) w^T\\
		&= -2r w^T\label{eq:partial-derivatives}
\end{align}
where $r:=\Delta_{\Omega_t} (v_t -  U w)$ denotes the (zero padded) residual vector and $w$ is the least-squares solution in~\eq{f-def}.   

Using Equation (2.70) in~\cite{Edelman98}, we can calculate the gradient on the Grassmannian from this partial derivative
\begin{equation*}
\begin{aligned}
	\nabla F &= (I - UU^T) \frac{dF}{dU}\\
	& =  -2  (I - UU^T) r w^T= -2 r w^T\,.
\end{aligned}
\end{equation*}
The final equality follows because the residual vector $r$ is orthogonal to all of the columns of $U$. This can be verified from the definitions of $r$ and $w$. 

A gradient step along the geodesic with tangent vector $-\nabla F$ is given by Equation (2.65) in~\cite{Edelman98}, and is a function of the singular values and vectors of $\nabla F$. It is trivial to compute the singular value decomposition of $\nabla F$, as it is rank one.  The sole non-zero singular value is $\sigma = 2||r|| ||w||$ and the corresponding left and right singular vectors are $\tfrac{r}{\|r\|}$ and $\frac{w}{\|w\|}$ respectively. Let $x_2,\ldots, x_{d}$ be an orthonormal set orthogonal to $r$ and $y_2,\ldots, y_d$ be an orthonormal set orthogonal to $w$. Then
\begin{equation*}
\begin{aligned}
-2 r w^T  = \left[\begin{array}{cccc} -\frac{r}{\|r\|} & x_2& \ldots & x_d \end{array}\right] \times \diag(\sigma,0,\ldots,0) \times \left[\begin{array}{cccc} \frac{w}{\|w\|} & y_2& \ldots & y_d \end{array}\right]^T
\end{aligned}
\end{equation*}
forms an SVD for the gradient. Now using (2.65) from~\cite{Edelman98}, we find that for $\eta>0$, a step of length $\eta$ in the direction $\nabla F$ is given by
\begin{align*}
U(\eta) &= U + \frac{(\cos(\sigma\eta)-1)}{\|w\|^2} Uww^T + \sin(\sigma \eta) \frac{r}{\|r\|} \frac{w^T}{\|w\|}\\
&= U + \left(  \sin(\sigma \eta)  \frac{r}{\|r\|}+ (\cos(\sigma \eta)-1) \frac{p}{\|p\|}  \right)\frac{w^T}{\|w\|}
\end{align*}
where $p:= Uw$, the predicted value of the projection of the vector $v$ onto $S$.  

This geodesic update rule is remarkable for a number of reasons.  First of all, it consists only of a rank-one modification of the current subspace basis $U$. Second, the term $\frac{\sin(\sigma \eta)}{\|r\|\|w\|} = \frac{\sin(\sigma \eta)}{\sigma}$ is on the order of $\eta$ when $\sigma\eta$ is small.   That is, for small values of $\sigma$ and $\eta$ this expression looks like a normal step along the gradient direction $-2r w^T$ given by~\eq{partial-derivatives}.  From the Taylor series of the cosine, we see that the second term is approximately equal to $\sigma^2\eta^2 \frac{p w^T}{\|p\|\|w\|}$.  That is, this term serves as a second order correction to keep the iterates on the Grassmannian.  Surprisingly, this simple additive term maintains the orthogonality, obviating the need for orthogonalizing the columns of $U$ after a gradient step. Below, we will also discuss how this iterate relates to more familiar iterative algorithms from linear algebra which use full information.

The GROUSE algorithm simply follows geodesics along the gradients of $F$ with a prescribed set of step-sizes $\eta$. The full computation is summarized in Algorithm~\ref{alg:stochqr}.  Our derivations have shown that computing a gradient step only requires the solution of the least squares problem~\eq{f-def}, the computation of $p$ and $r$, and then a rank one update to the previous subspace.

Each step of GROUSE can be performed efficiently with standard linear algebra packages.  Computing the weights in Step~\ref{step:lsq} of Algorithm~\ref{alg:stochqr} requires solving a least squares problem in $|\Omega_t|$ equations and $d$ unknowns.  Such a system is solvable in at most
$O(|\Omega_t|d^2)$ flops in the worst case.  Predicting the component of $v$ that lies in the current subspace requires a matrix vector multiply that can be computed in $O(nd)$ flops.  Computing the residual then only requires $O(|\Omega_t|)$ flops, as we will always have zeros in the entries indexed by the complement of $\Omega_t$.  Computing the norms of $r$ and $p$ can be done in $O(n)$ flops.  The final subspace update consists of adding a rank one matrix to an $n \times d$ matrix and can be computed in $O(nd)$ flops.  Totaling all of these computation times gives an overall complexity estimate of $O(nd+|\Omega_t|d^2)$ flops per subspace update.

\begin{algorithm}  \caption{Grassmannian Rank-One Update Subspace Estimation}
  \begin{algorithmic}[1]
    \REQUIRE An $n \times d$ orthogonal matrix $U_0$. A sequence of vectors $v_t$, each observed in entries $\Omega_t$. A set of stepsizes $\eta_t$.
    \FOR{$t=1,\ldots,T$}
    \STATE {\bf Estimate weights:} $w = \arg\min_a \|\Delta_{\Omega_t} (U_t a - v_t)\|^2$\label{step:lsq}
    \STATE {\bf Predict full vector:} $p = U_t w$ \label{step:parallel}
    \STATE {\bf Compute residual:} $r = \Delta_{\Omega_t} (v_t-p)$\label{step:res}
    \STATE {\bf Update subpace}:  $U_{t+1} = U_t+ \left(  (\cos(\sigma \eta_t)-1) \frac{p}{\|p\|} + \sin(\sigma \eta_t)  \frac{r}{\|r\|} \right)\frac{w^T}{\|w\|}$ \\ 
 \qquad  \qquad \qquad  \qquad   where $\sigma=\|r\|\|p\|$    \label{step:update}
 \ENDFOR
  \end{algorithmic}
  \label{alg:stochqr}
\end{algorithm}

\subsection{Step-sizes}

In the static case where the subspace $S[t] = S_0$ for all $t$, we can guarantee that the algorithm converges to a stationary point of~\eq{global-opt} as long as the stepsizes $\eta_t$ satisfy
\begin{equation*}
	\lim_{k\rightarrow\infty}\eta_t = 0 \quad \mbox{ and } \quad \sum_{k=1}^\infty \eta_t = \infty
\end{equation*}
Selecting $\eta_t\propto 1/t$ will satisfy this assumption.  This analysis appeals to the classical ODE method~\cite{KushnerBook}, and we are guaranteed such convergence because $\mathfrak{G}(n,d)$ is compact.   

In the case that $S[t]$ is changing over time, a constant stepsize  is needed to continually adapt to the changing subspace. Of course, if a non-vanishing stepsize is used then the error will not converge to zero, even in the static case.  This leads to the common tradeoff between tracking rate and steady-state error in adaptive filtering problems. We explore this tradeoff in Section~\ref{sec:experiments}.


\subsection{Comparison to methods that use full information}

GROUSE can be understood as an adaptation of an incremental update to a QR or SVD factorization.   Most batch subspace identification algorithms that rely on the eigenvalue decomposition, the singular value decomposition, or their more efficient counterparts such as the QR decomposition or the Lanczos method, can be adapted for on-line updates and tracking of the principal subspace.  A comprehensive survey of these methods can be found in~\cite{comon90}.

Suppose we fully observe the vector $v$ at each increment.  Given an estimated basis, $U_\mathrm{est}$ for the unknown subspace $S$,  we would update our estimate for $S$ by computing the component of $v$ that is orthogonal to $U$.  We would then append this new orthogonal component to our basis $U$, and use an update rule based on the magnitude of the component of $v$ that does not lie in the span of $U$.  Brand~\cite{Brand06} has shown that this method can be used to compute an efficient incremental update of an SVD using optimized algorithms for computing rank one modifications of matrix factorizations~\cite{Gu94}. 

In lieu of being able to exactly able to compute component of $v_t$ that is orthogonal to our current subspace estimate, GROUSE computes this component only on the entries in $\Omega_t$. Recent work~\cite{Balzano10} shows that for generic subspaces, the estimate for $r$ computed by Algorithm~\ref{alg:stochqr} in a single iteration is an excellent proxy for the amount of energy that lies in the subspace, provided that the number of measurements at each time step is greater than the true subspace dimension times a logarithmic factor.  One can also verify that the GROUSE update rule corresponds to forming the matrix $[U,r]$ and then truncating the last column of the matrix
\begin{equation*}
[U,r] R_\eta
\end{equation*}
where $R_\eta$ denotes the $(r+1) \times (r+1)$ rotation matrix
\begin{equation*}
	R_\eta = \left[\begin{array}{cc} I - \frac{ww'}{\|w\|}(1-\cos(\eta \sigma)) & -\frac{w}{\|w\|} \sin(\eta\sigma)\\ \frac{w}{\|w\|} \sin(\eta\sigma)  & \cos(\eta \sigma) \end{array}\right]\,.
\end{equation*}
That is, our algorithm computes a mixture of the current subspace estimate and a predicted orthogonal component. This mixture is determined both by the stepsize and the relative energy of $v_t$ outside the current subspace.

\section{Numerical Experiments}\label{sec:experiments}

\subsection{Subspace Identification and Tracking}

\paragraph{Static Subspaces} We first consider the problem of identifying a fixed subspace.  In all of the following experiments, the full data dimension is $n=700$, the rank of the underlying subspace is $d=10$, and the sampling density is $0.17$ unless otherwise noted.
We generated a series of iid vectors $v_t$ according to generative model:
\begin{equation*}
	v_t = U_{\mathrm{true}} \alpha + \beta
\end{equation*}
where $U_{\mathrm{true}}$ is an $n\times d$ matrix whose $d$ orthonormal columns spam the subspace, $\alpha$  is a $d\times 1$ vector whose entries are realizations of iid ${\cal N}(0,1)$ random variables, and $\beta$ is an $n\times 1$ vector whose entries
are iid ${\cal N}(0,\omega^2)$. Here ${\cal N}(0,\omega^2)$ denotes the Gaussian distribution with mean zero and variance $\omega^2 >0$ which governs the SNR of our data.

\begin{figure*}[htb]
\centering
\begin{tabular}{cc}
  \includegraphics[width=5cm]{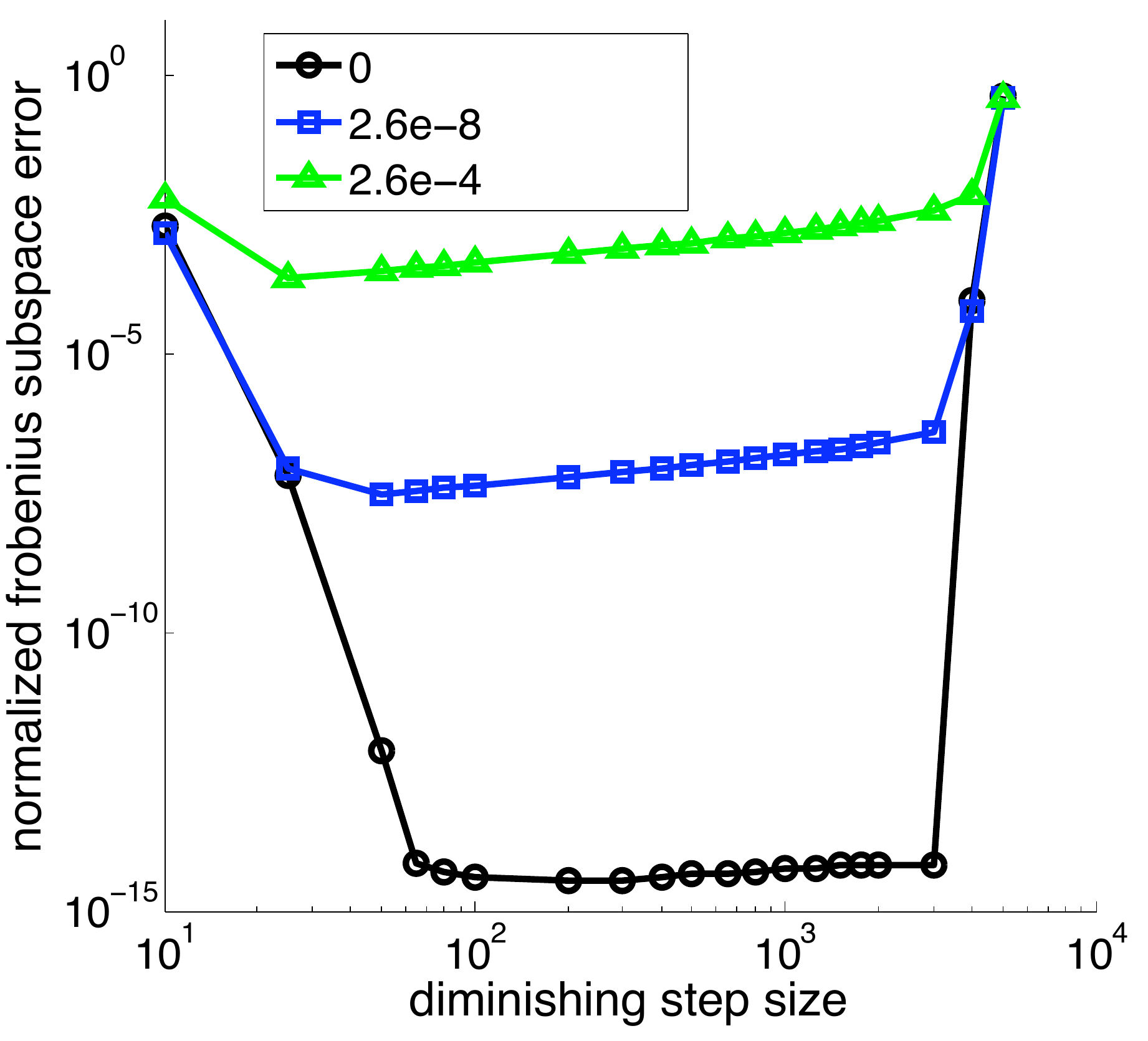} &
  \includegraphics[width=5cm]{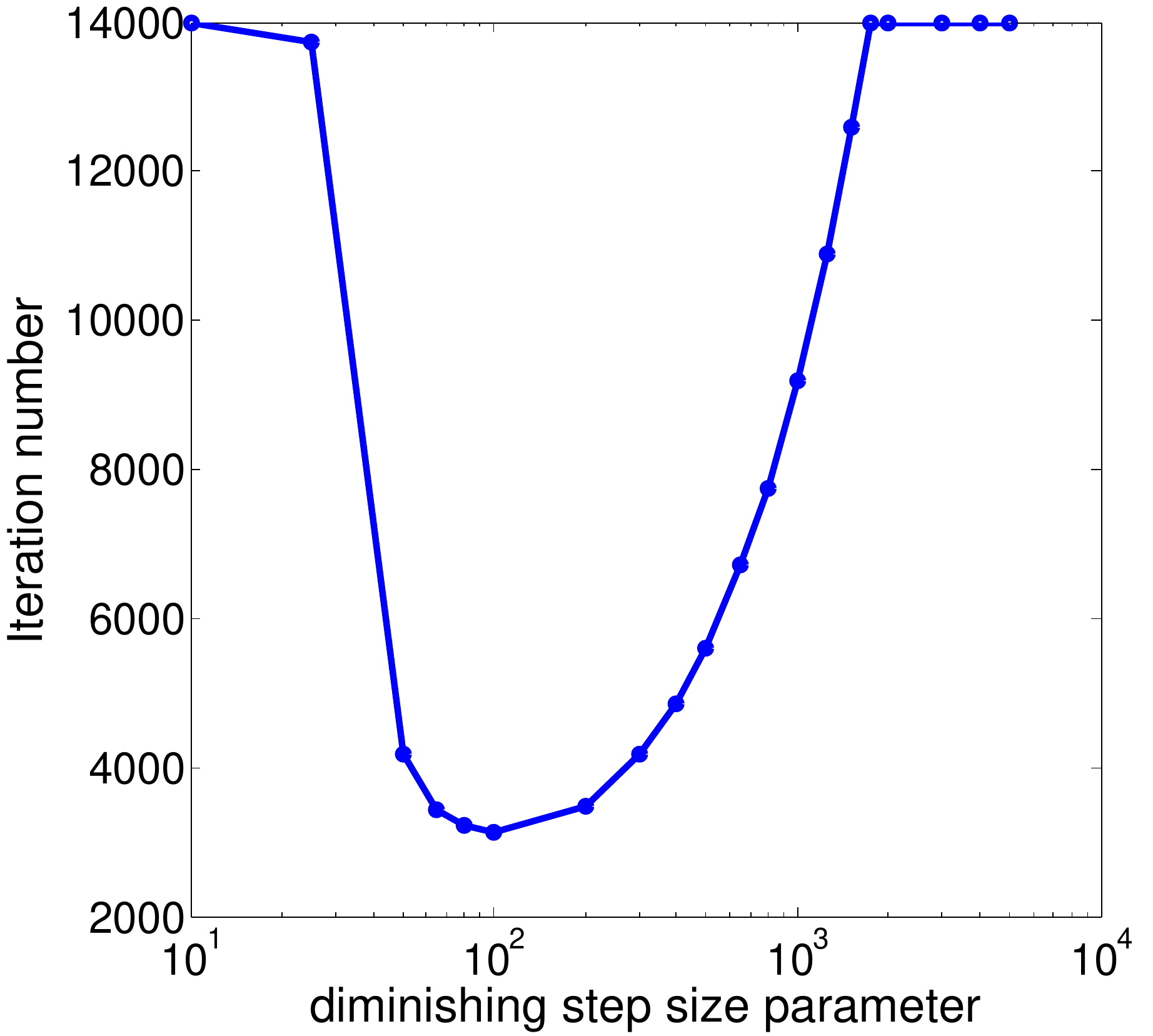}\\
  (a) & (b)\\
  \includegraphics[width=5cm]{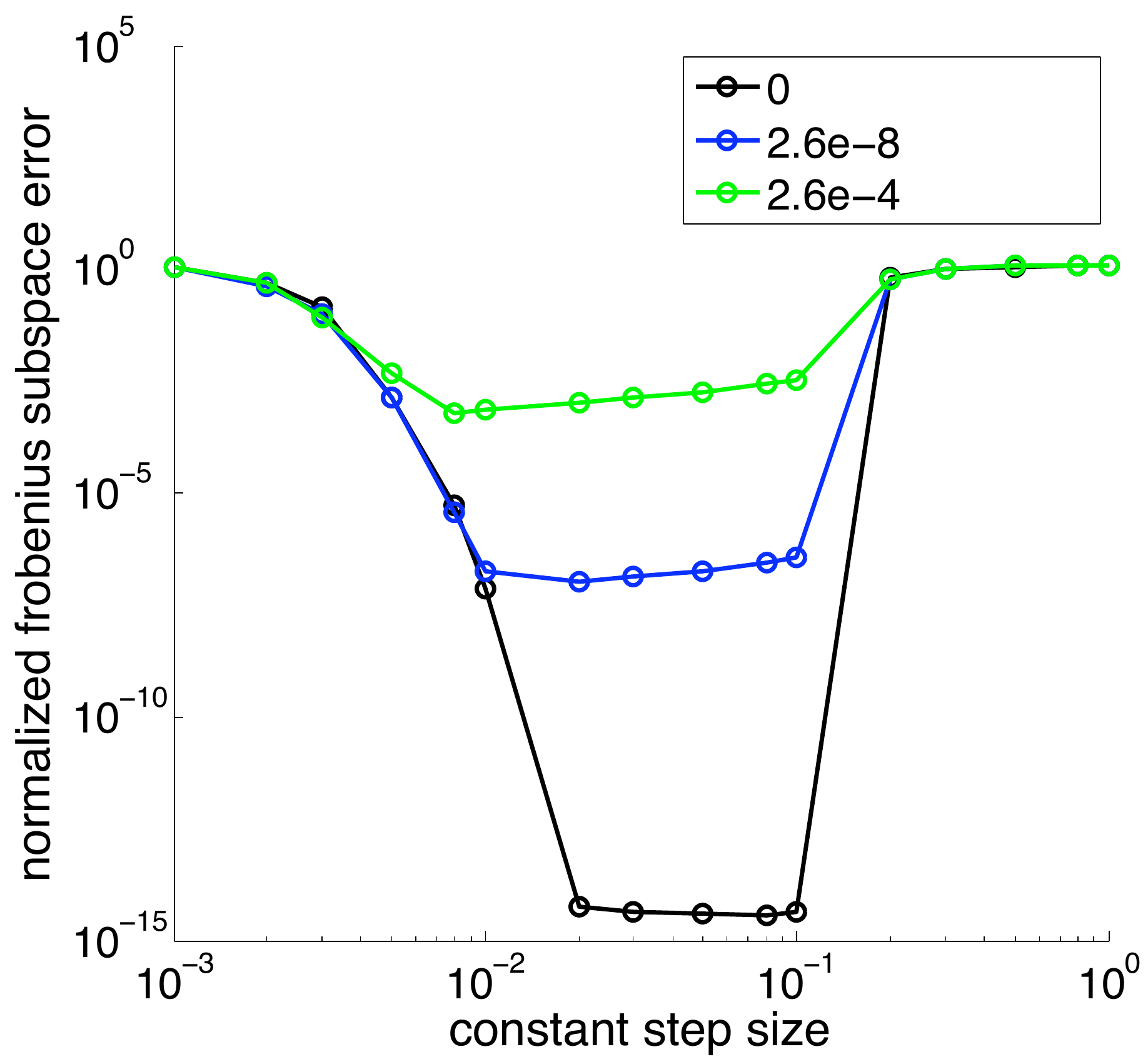}&
  \includegraphics[width=5cm]{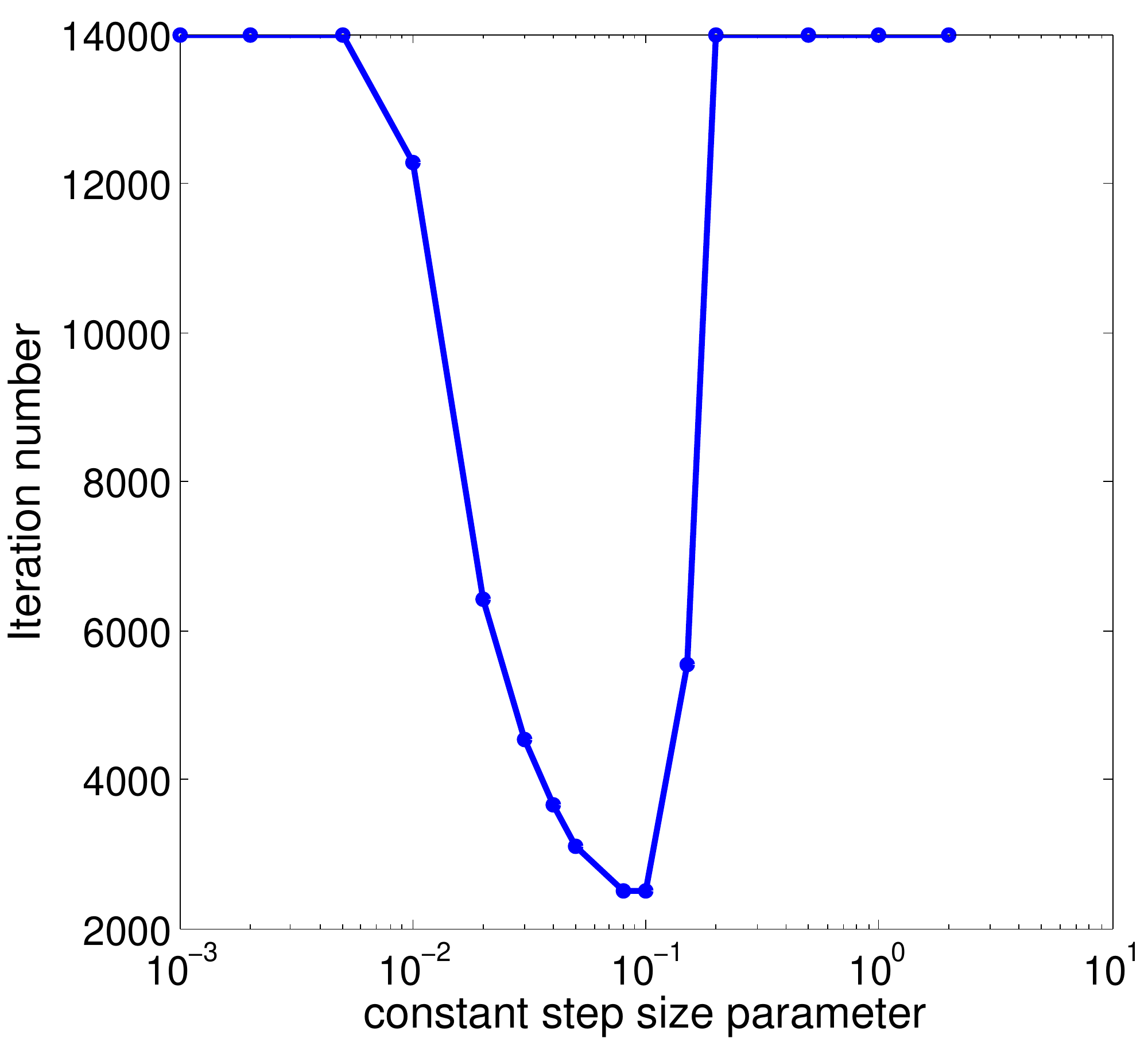}\\
  (c) & (d)
  \end{tabular}
    \caption{\small (a) Performance sensitivity to noise for a diminishing stepsize policy $\eta_t=C/t$. Results are displayed for three values of the noise magnitude, $\omega$. (b) The number of iterations required to achieve an error of $10^{-6}$ as a function of $C$. (c) and (d) are the same except that the stepsize policy is here $\eta_t = C$.}
    \label{fig:noise}
\end{figure*}

We implemented GROUSE (see Algorithm~\ref{alg:stochqr} above) with a stepsize rule of $\eta_t:=C/t$ for some constant $C>0$. Figure~\ref{fig:noise}(a) shows the steady state error of the tracked subspace with varying values of $C$ and the noise variance.  All the data points reflect the error performance at $t=14000$. We see that GROUSE  converges for $C$ ranging over an order of magnitude, however with additive noise the smaller stepsizes yield smaller errors. When there is no noise, i.e., $\omega^2=0$, the error performance is near the level of machine precision and is flat for the whole range of converging stepsizes.
In Figure~\ref{fig:noise}(b) we show the number of input vectors after which the algorithm converges to an error of less than $10^{-6}$. The results are consistent with Figure~\ref{fig:noise}(a), demonstrating that smaller stepsizes in the suitable range take fewer vectors until convergence. We only ran the algorithm up to time $t=14000$, so the data points for the smallest and the largest few stepsizes only reflect that the algorithm did not yet reach the desired error in the allotted time.
Figures~\ref{fig:noise}(c) and~\ref{fig:noise}(d) repeat identical experiments with a constant stepsize policy, $\eta_t = C$.  We again see a wide range of stepsizes for which GROUSE converges, though the region of stability is narrower in this case.

\begin{figure*}[htb]
\centering
\begin{tabular}{ccc}
 \includegraphics[width=5cm]{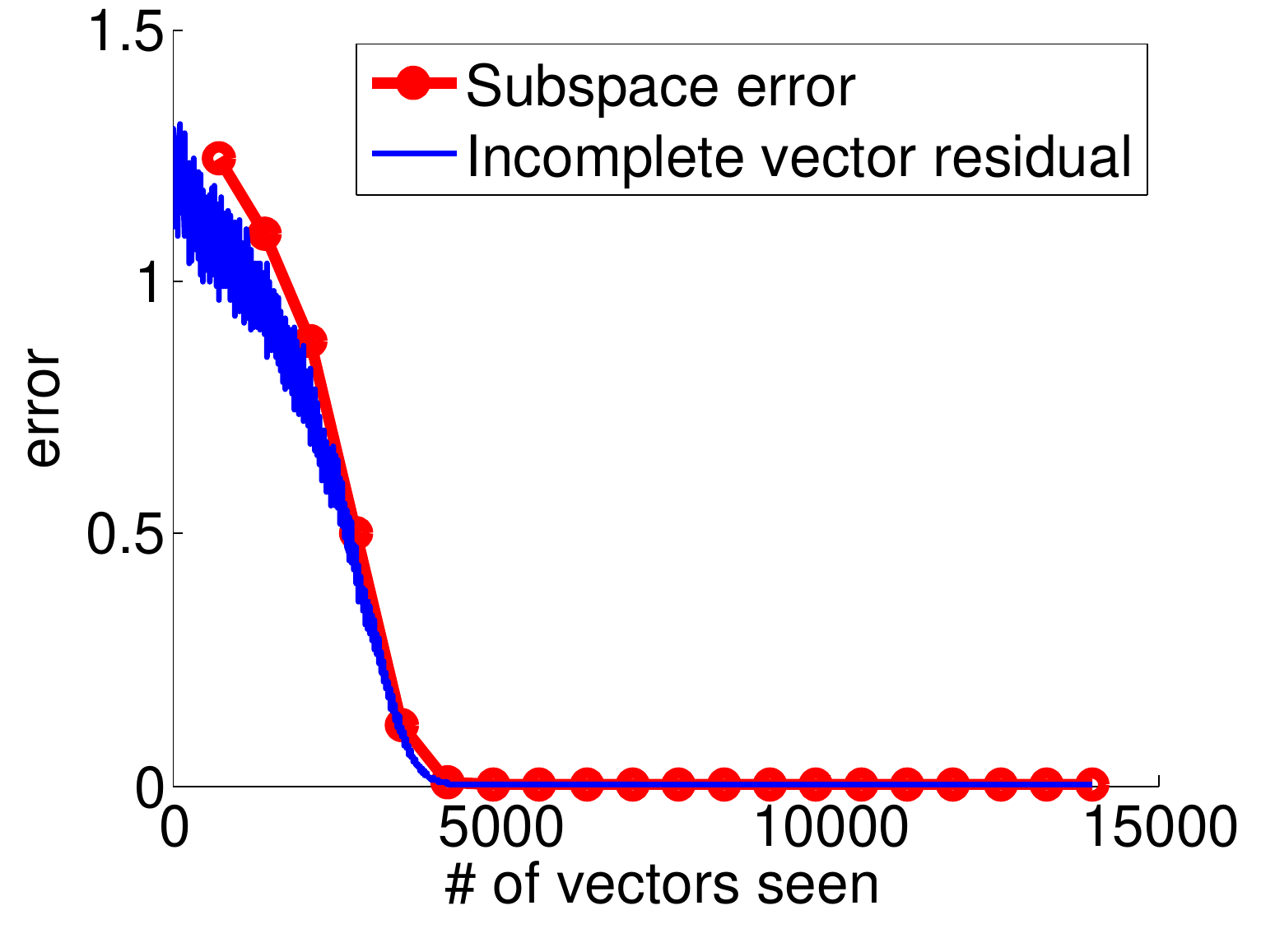} &
 \includegraphics[width=5cm]{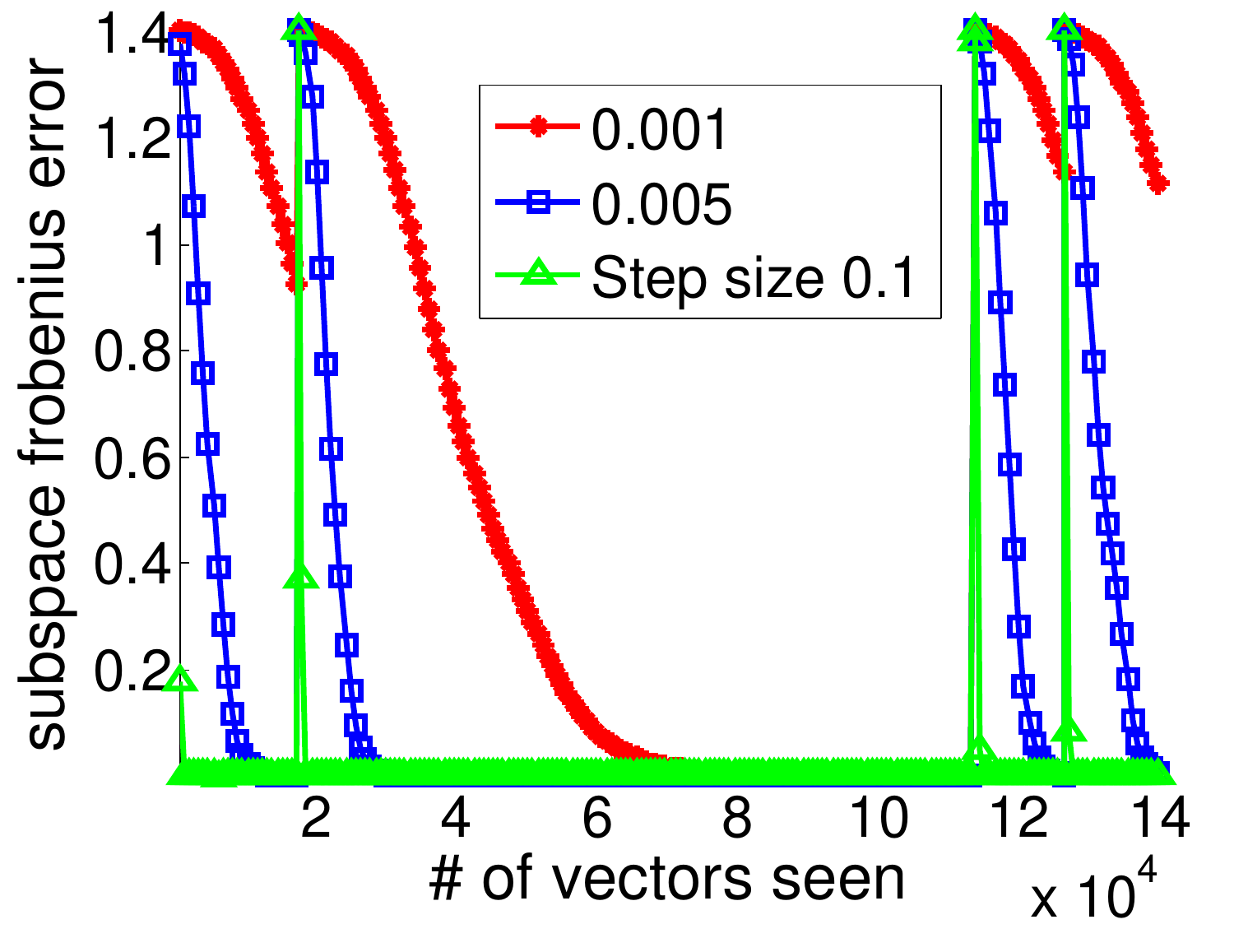}  &
  \includegraphics[width=5cm]{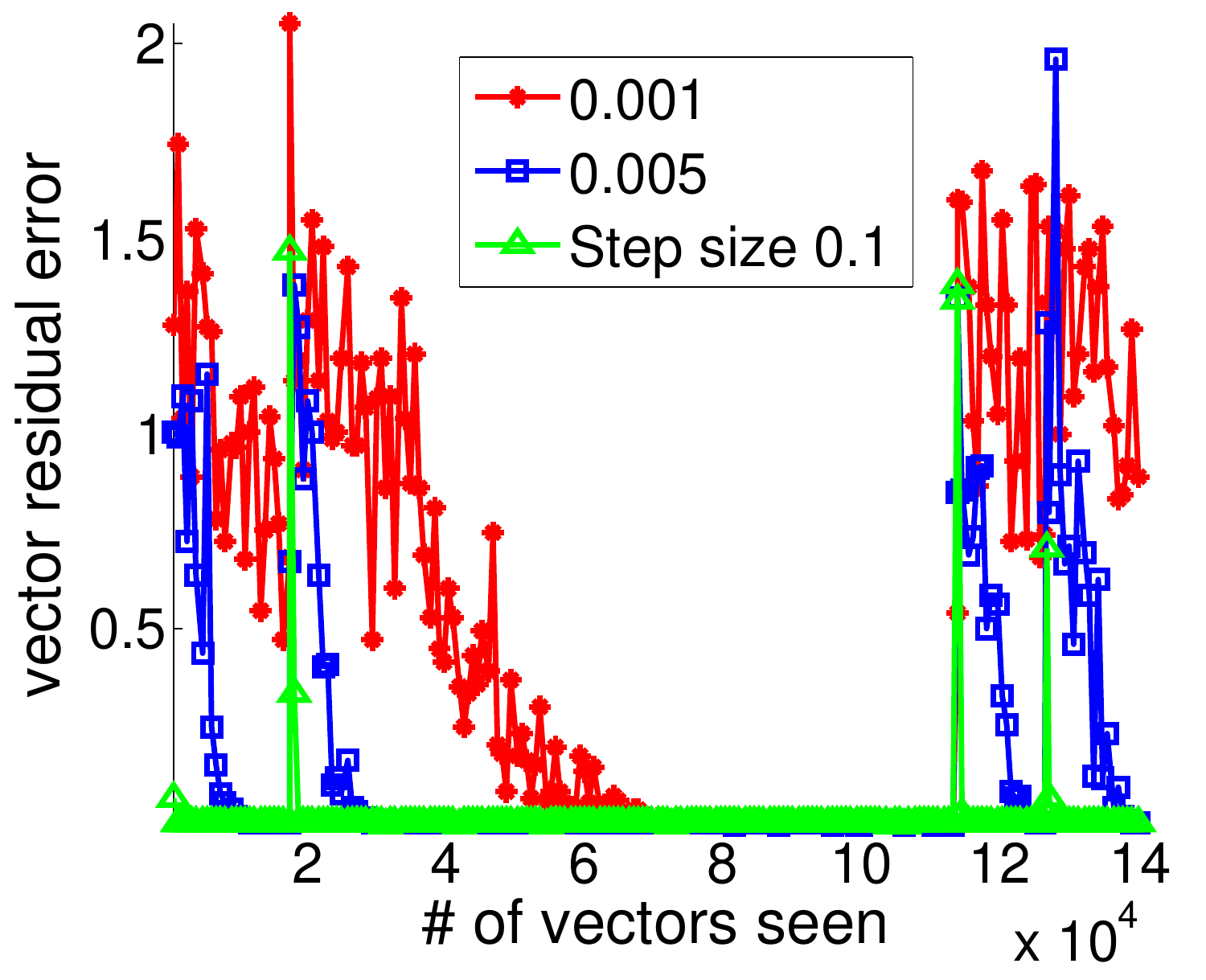} \\
  (a) & (b) & (c)
  \end{tabular}
   \caption{\small (a) Comparison of the distance to the true subspace and the norm of the residual in Step~\ref{step:res} of the GROUSE algorithm.  The residual norm closely tracks the distance to the actual subspace. (b) Using constant stepsize to track sudden changes in subspace.  We plot the transient behavior three constant stepsize policies.  In (c), we again verify that the norm of the residual gives an accurate signature for anomaly detection and for tracking success.}
    \label{fig:suddenchange}
\end{figure*}

We note that the norm of the residual $\|r\|$ provides an excellent indicator for whether tracking is successful. 
A shown in Figure~\ref{fig:suddenchange}(a), the error to the true subspace is closely approximated by $\|r\|/\|v_t\|$.  This confirms the theoretical analysis in~\cite{Balzano10} which proves that this residual norm is an accurate estimator of the true subspace error when the number of samples is appropriately large.

%


\paragraph{Subspace Change Detection}

As a first example of GROUSE's ability to adapt to changes in the underlying subspace, we simulated a scenario where the underlying subspace abruptly changes at three points over the course of an experiment with 14000 observations.  At each break, we selected a new subspace $S$ uniformly at random and GROUSE was implemented with  a constant stepsize. As is to be expected, the algorithm is able to re-estimate the new subspace in a time depending on the magnitude of the constant stepsize.

\paragraph{Rotating Subspace}

\begin{figure*}[htb]
\centering
\begin{tabular}{cccc}
  \includegraphics[width=3.5cm]{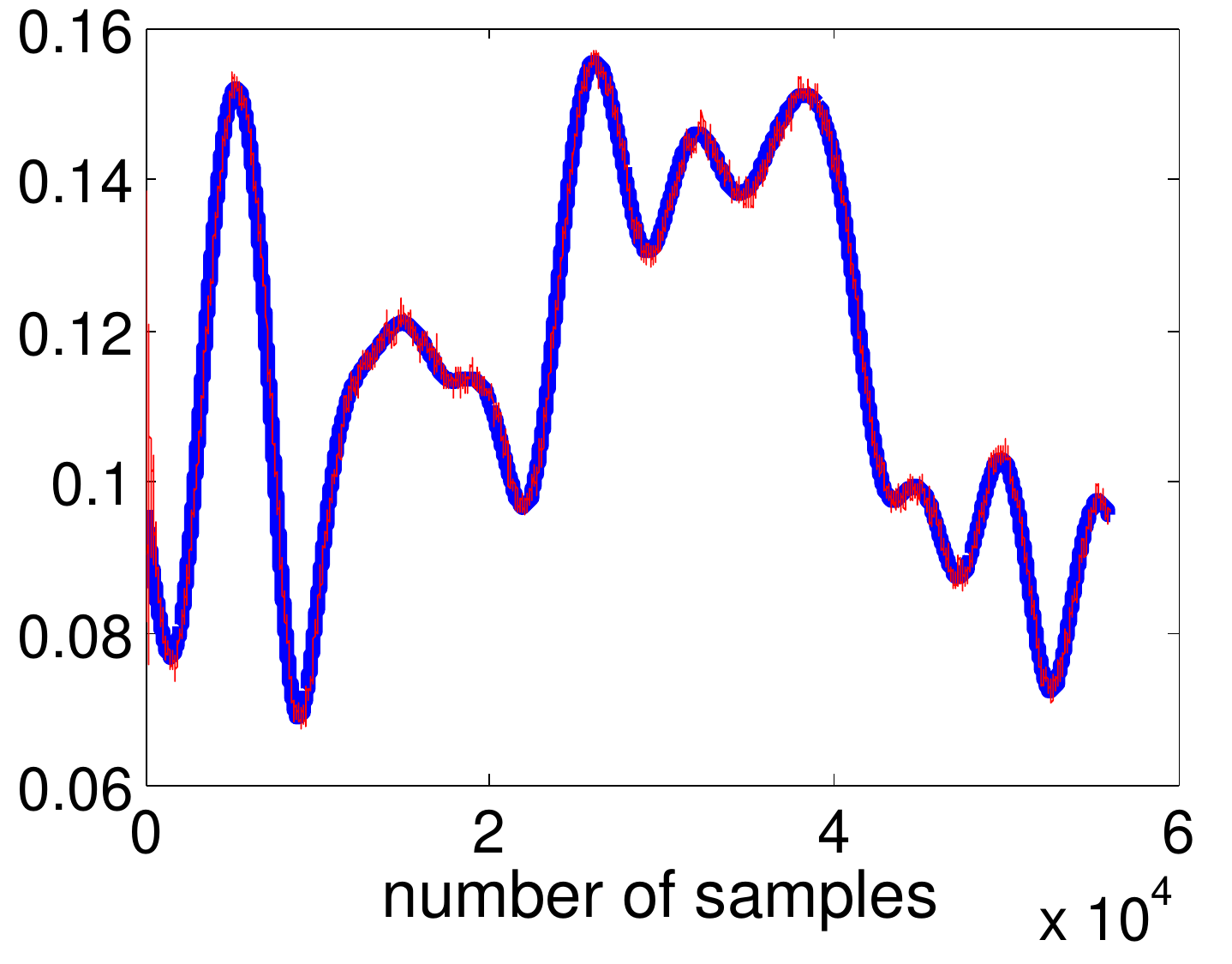} &
    \includegraphics[width=3.5cm]{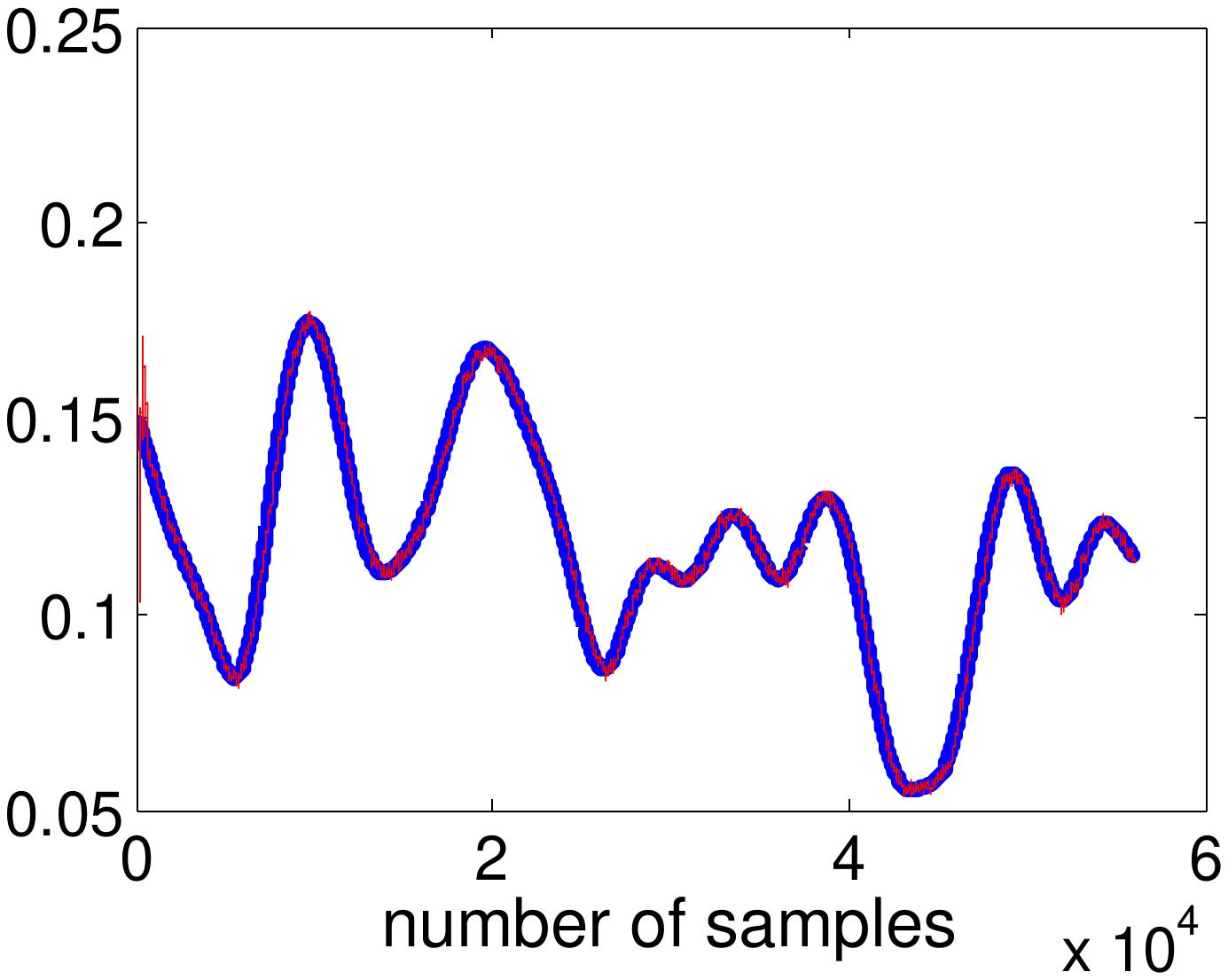} &
      \includegraphics[width=3.5cm]{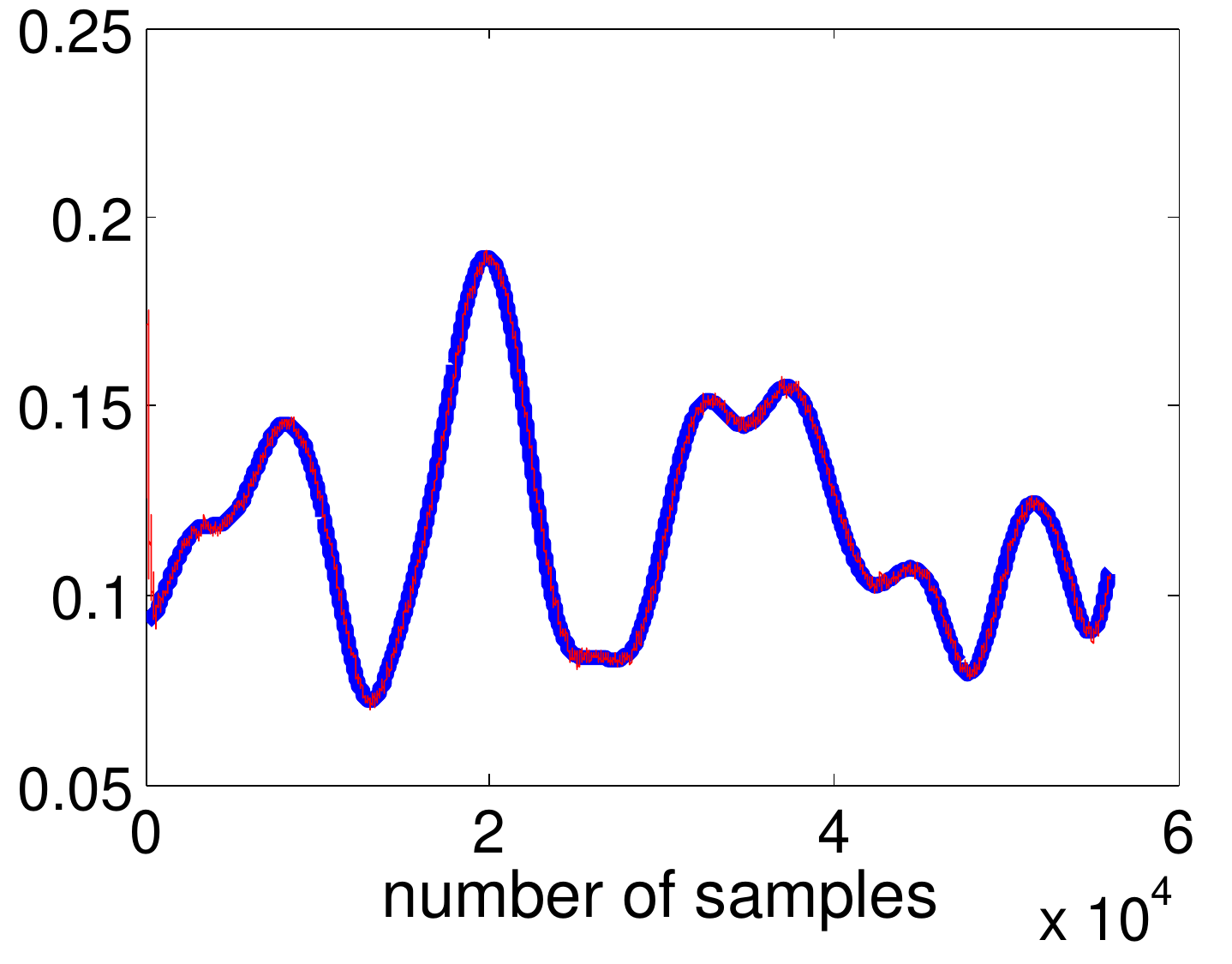} &
        \includegraphics[width=3.5cm]{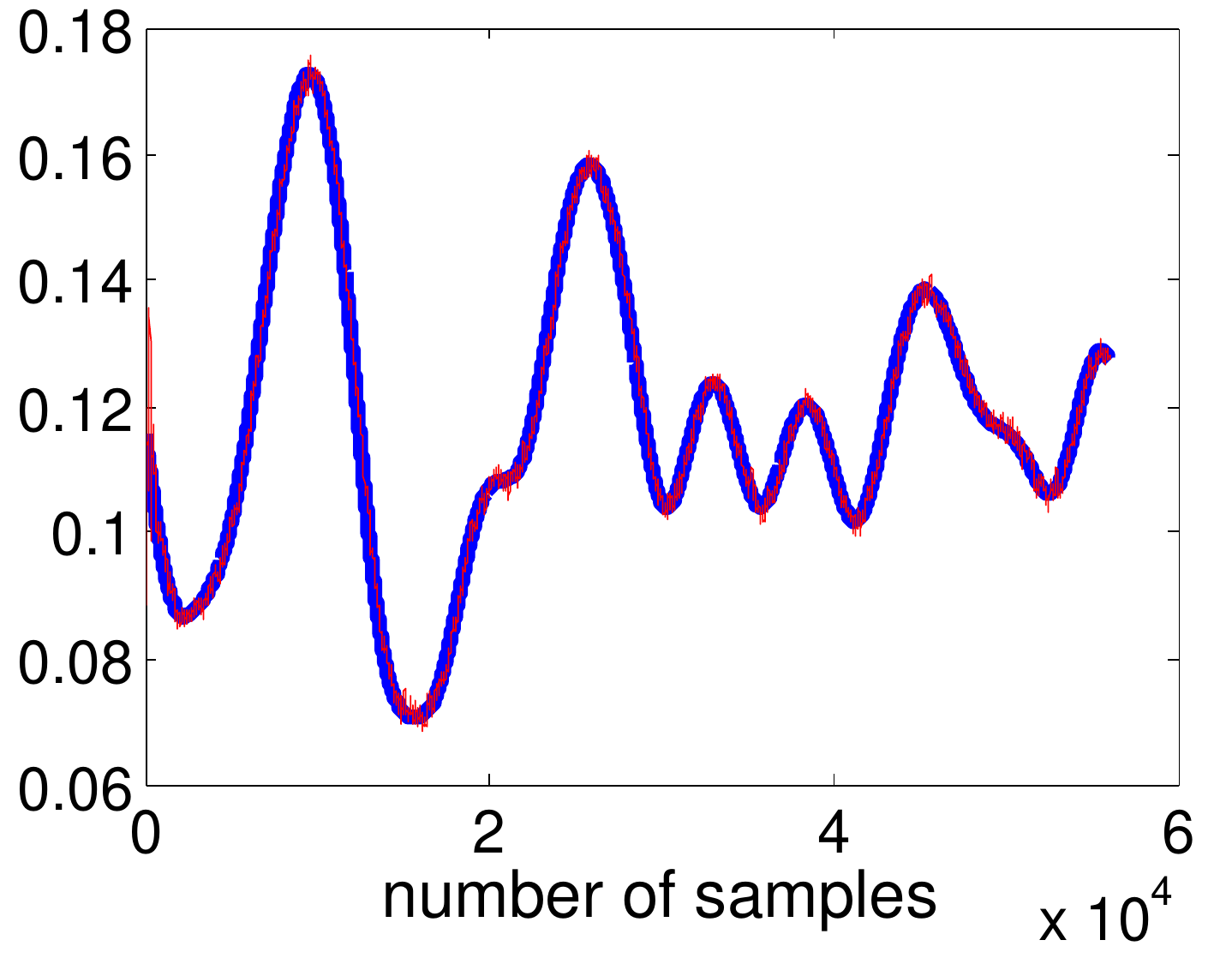} 
  \end{tabular}
    \caption{\small Tracking a rotating subspace.  Here we plot the norm of the projection of four random vectors over time.  The blue curves denote the true values of these norms, and the red curves plot the GROUSE predictions.  Note that except for very early transients, GROUSE fully tracks the subspace. }
        \label{fig:rotation}
\end{figure*}

In this second synthetic experiment, the subspace evolves according to a random ordinary differential equation.  Specifically, we sample a skew-symmetric matrix $B$ with independent, normally distributed entries and set
\begin{equation*}
	\dot{U} = B U\,,\quad U[0] = U_0.
\end{equation*}
The resulting subspace at each iteration is thus $U[t] = \exp(\delta t B)$ where $\delta$ is a positive constant. The resulting subspace at each iteration is thus $U[t] = \exp(\delta t B)$ where $\delta$ is a positive constant. In Figure~\ref{fig:rotation}, we show the results of tracking the rotating subspace with $\delta =10^{-5}$. To demonstrate the effectiveness of the tracking, we display the projection of  four random vectors using both the true subspace (in blue) and our subspace estimate at that time instant (in red).

\paragraph{Tracking Chlorine Levels}

\begin{figure*}
\centering
\hspace{-1.25in}\begin{minipage}[c]{5in}
\centering
\begin{tabular}{cc}
\includegraphics[width=5cm]{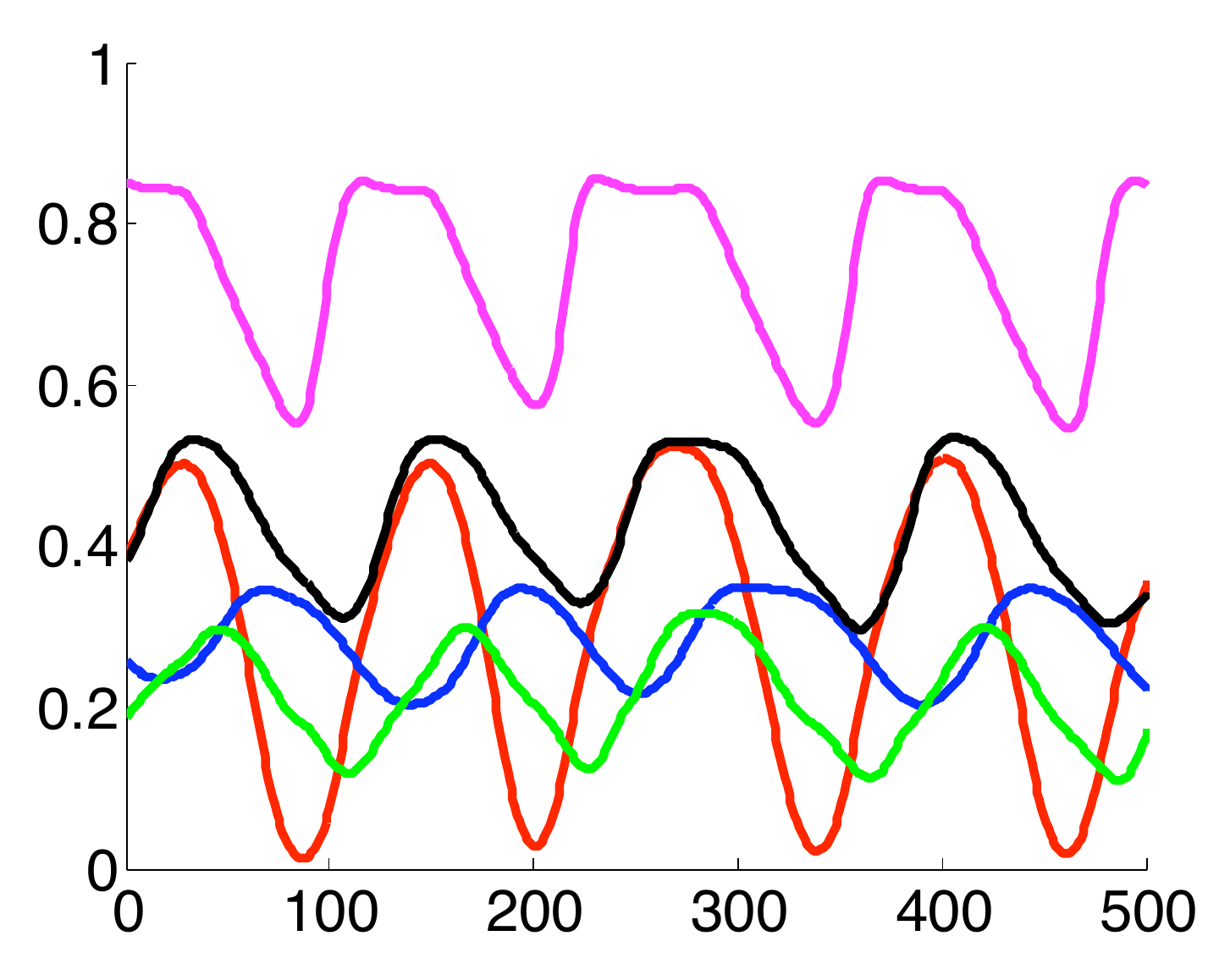}&
\includegraphics[width=5cm]{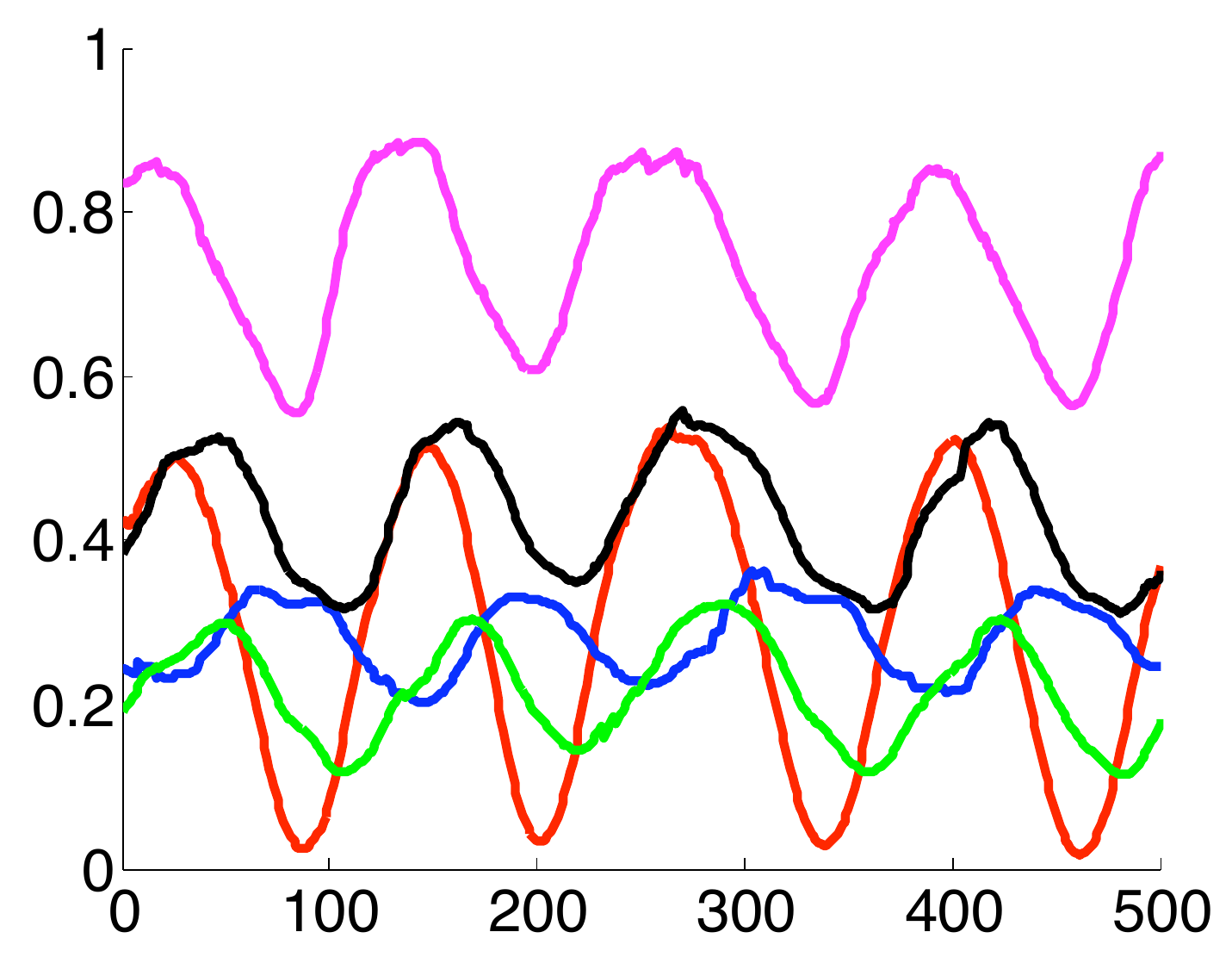}\\
  (a) & (b)
  \end{tabular}
\end{minipage}
\begin{minipage}[c]{1in}
\scriptsize 
\begin{center}
\begin{tabular}{c|c|c}
\hline
 &  &  reconstruction\\
$|\Omega_t|/n$ & $\eta_t$ &  error\\
\hline
\hline
	 0.2  &   5e-3 &  0.2530 \\
	  & 3e-2 &     0.1244   \\
	\hline
	 0.4 & 5e-3 &        0.1797  \\
	 &    1e-2 &  0.1233  \\
	 &    3e-2 &   0.1432 \\
	 \hline
	 0.7 &   5e-3 &   0.1289  \\
	 &   7e-3 &    0.1221  \\
	 &   3e-2 &     0.1684 \\
	 \hline
	  1 &  5e-3 &    0.1253 \\
	   &     3e-2 &  0.2217  \\
	\hline 
  \end{tabular}
\end{center}
\end{minipage}
\caption{\small \label{fig:chlorine}
 (a) Actual sensor readings. (b) Predicted sensor readings from tracked subspace.  In these figures we are displaying the values at sensors 4, 17, 148, 158, and 159.  The table lists errors in tracking the chlorine data set for varying sampling densities and stepsizes.  The error between the data and the best SVD approximation is 0.0704.
}
\end{figure*}

We also analyzed the performance of the GROUSE algorithm on simulated chlorine level monitoring in a pressurized water delivery system.  The data were generated using EPANET~\footnote{\url{http://www.epa.gov/nrmrl/wswrd/dw/epanet.html}} software and were previously analyzed ~\cite{papadimitriou05}. The input to EPANET is a network layout of water pipes and the output has many variables including the chemical levels, one of which is the chlorine level. The data we used is available from ~\cite{papadimitriou05}~\footnote{\url{http://www.cs.cmu.edu/afs/cs/project/spirit-1/www/}}. 
This dataset has ambient dimension $n=166$, and $T=4610$ data vectors were collected, once every 5 minutes over 15 days. We tracked an $d=6$ dimensional subspace and compare this with the best 6-dimensional SVD approximation of the entire complete dataset.  The results are displayed in Figure~\ref{fig:chlorine}. The table gives the results for various constant stepsizes and various fractions of sampled data. The smallest sampling fraction we used was $20\%$, and for that the best stepsize was 3e-2; we also ran GROUSE on the full data, whose best stepsize was 5e-3. As we can see, the performance error improves for the smaller stepsize of 5e-3 as the sampling fraction increases; Also the performance error improves for the larger stepsize of 3e-2 as the sampling fraction decreases. However for all intermediate sampling fractions there are intermediate stepsizes that perform near the best reconstruction error of about 0.12. The normalized error of the full data to the best 3-dimensional SVD approximation is 0.0704. Note that we only allow for one pass over the data set and yet attain, even with very sparse sampling, comparable accuracy to a full batch SVD which has access to all of the data.

Figure~\ref{fig:chlorine}(a) and (b) show the original and the GROUSE reconstructions of five of the chlorine sensor outputs. We plot the last 500 of the 4310 samples, each reconstructed by the estimated subspace at that time instant.

\subsection{Matrix Completion Problems}
As described in Section~\ref{sec:mc}, matrix completion can be thought of as a subspace identification problem where we aim to identify the column space of the unknown low-rank matrix. Once the column space is identified, we can project the incomplete columns onto that subspace in order to complete the matrix. We have examined GROUSE in this context with excellent results. Our approach is to do descent on the column vectors in random order, and allow the algorithm to pass over those same incomplete columns a few times.

\begin{figure}[htb]
\centering
\begin{tabular}{cc}
  \includegraphics[width=5cm]{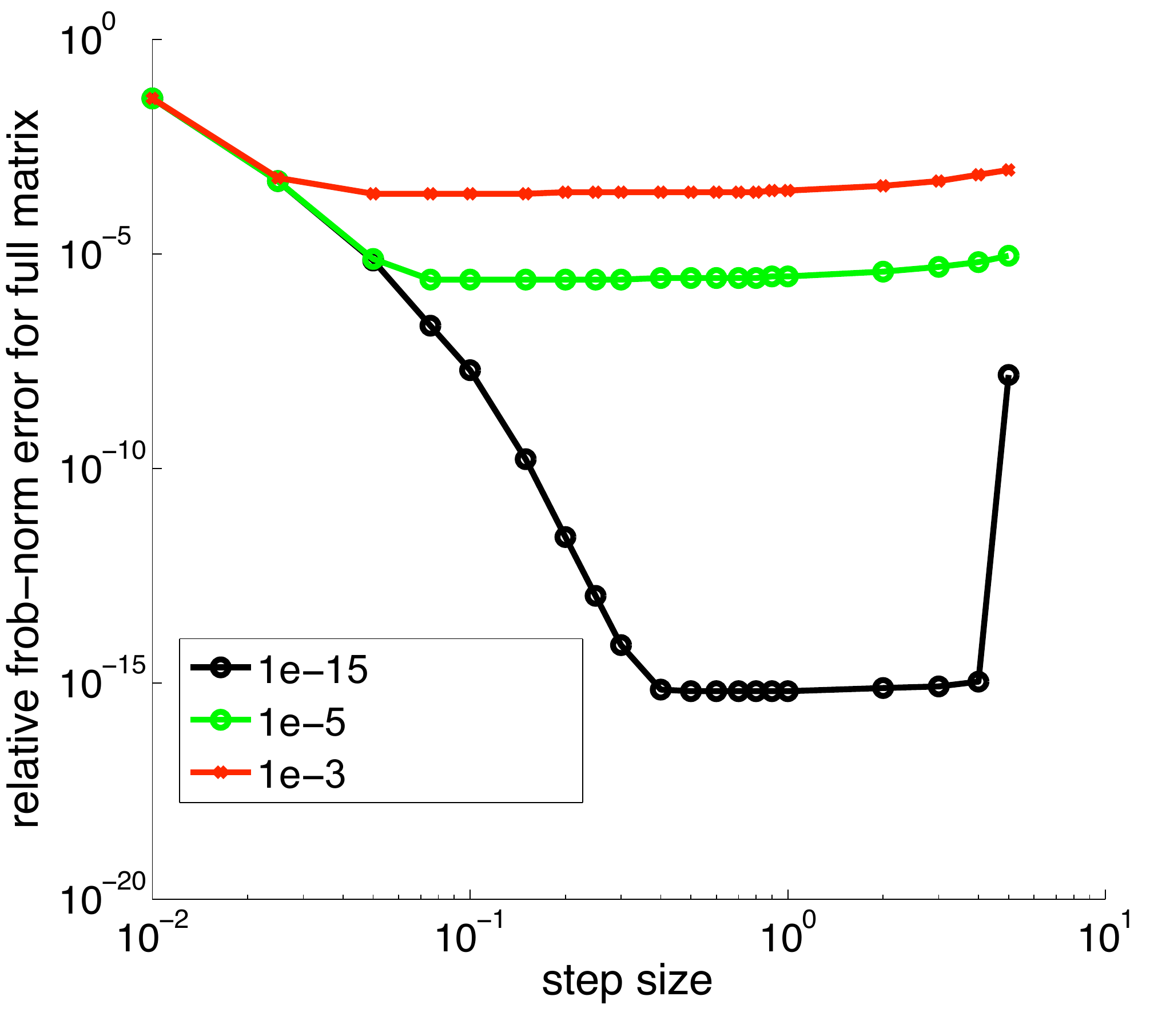} &
  \includegraphics[width=5cm]{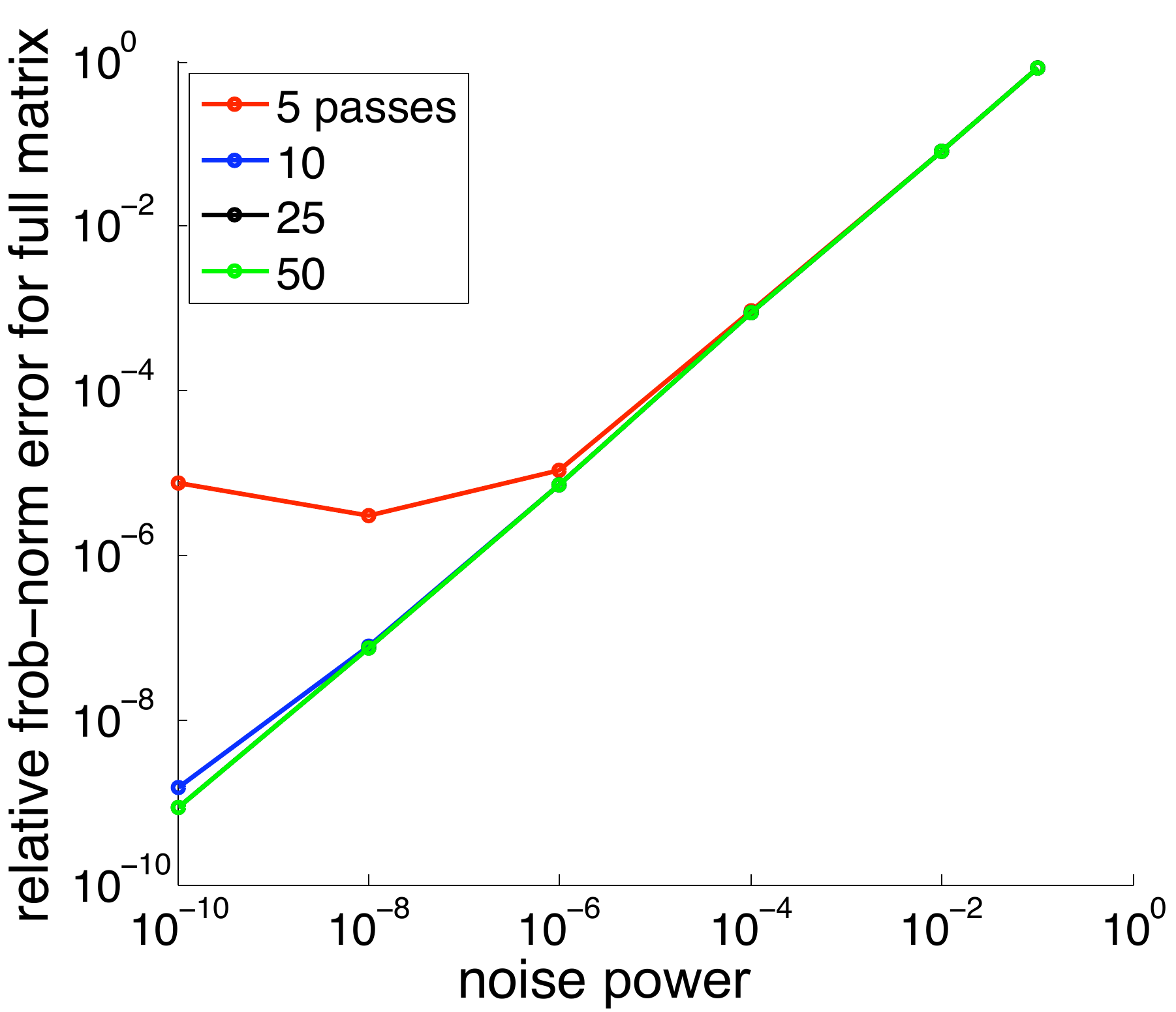} \\
(a) &  (b)
  \end{tabular}
    \caption{\small (a) Performance sensitivity to noise parameter $\omega$ and stepsize for matrix completion.  Here we use a diminishing stepsize $\eta = C/t$.  (b)  Here we plot the time to get to a desired Frobenius norm error on the hidden matrix.  We see that GROUSE converges to the noise floor after at most 10 passes over the columns of the matrix.}
    \label{fig:mc-performance}
\end{figure}

Our simulation set-up aimed to complete $700 \times 700$ dimensional matrices of rank $10$ sampled with density $0.17$.  We generated the low-rank matrix by generating two factors $Y_L$ and $Y_R$ with i.i.d. Gaussian entries, and added normally distributed noise with variance $\omega^2$.  The robustness to step-size and time to recovery are shown in Figure~\ref{fig:mc-performance}.

\begin{figure*}
\centering
\begin{minipage}[c]{2in}
\centering
  \includegraphics[width=2in]{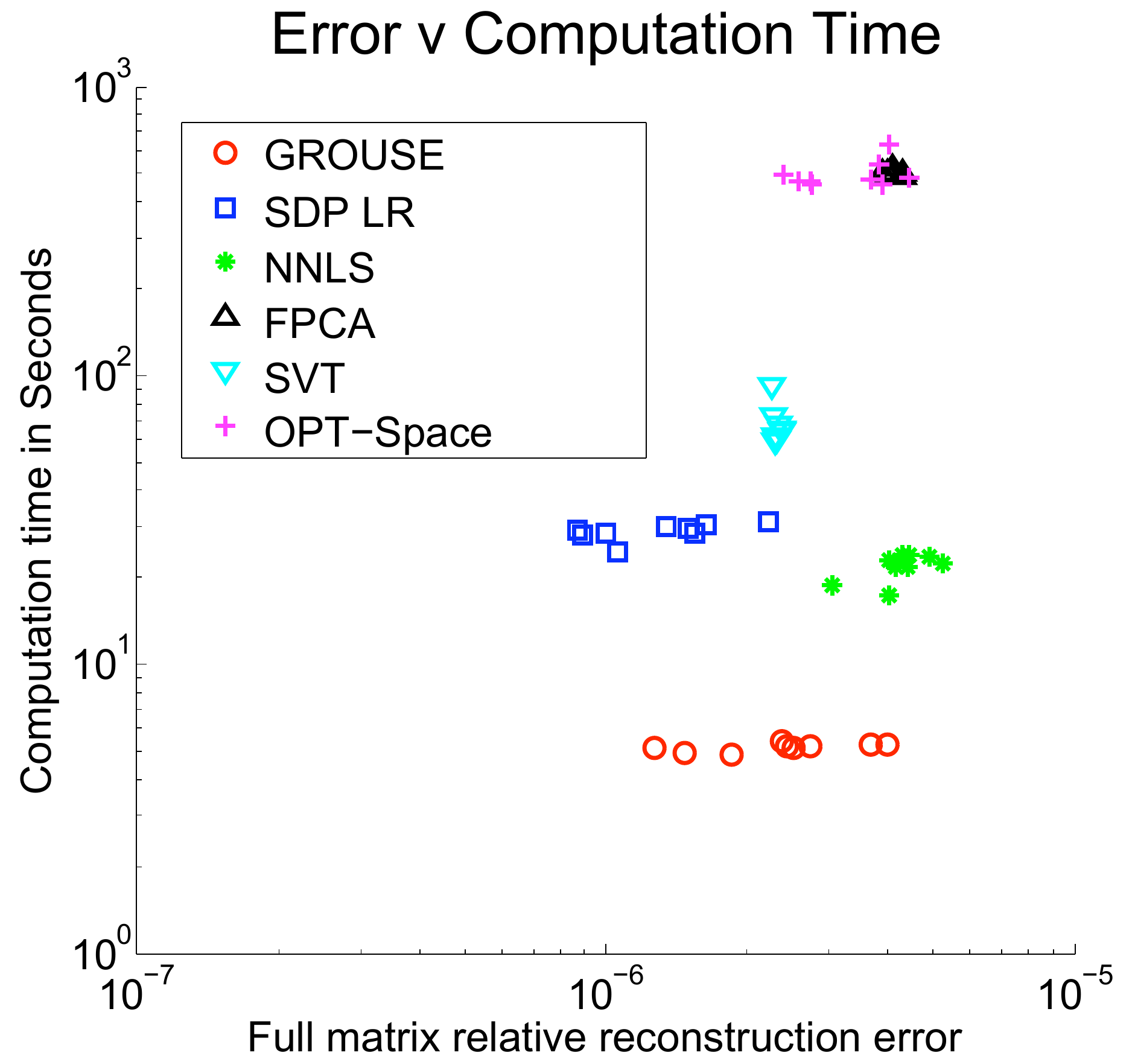}
\end{minipage}
  \begin{minipage}[c]{4in}
\begin{center}
\scriptsize 
\begin{tabular}{c c c c |c c c|c c}
\hline
\multicolumn{4}{c|}{Problem parameters} &
\multicolumn{3}{c|} {GROUSE}& 
\multicolumn{2}{c}{NNLS}\\
\hline
$n_r$ & $n_c$ & $r$  &dens & rel.err. & passes & time & rel. err. & time (s)\\
\hline
5000 & 20000 & 5 & 0.006 & 1.10e-4 & 2 & 14.8 & 4.12e-04 & 66.9 \\
5000 & 20000 & 10 & 0.012 & 1.5e-3 & 2 & 21.1 & 1.79e-4 & 81.2 \\
6000 & 18000 & 5 & 0.006 & 1.44e-5 & 3 & 21.3 & 3.17e-4 & 66.8 \\
6000 & 18000 & 10 & 0.011 & 8.24e-5 & 3 & 31.1 & 2.58e-4 & 68.8\\
7500 & 15000 & 5 & 0.005 & 5.71e-4 & 4 & 36.0 & 3.09e-4 & 62.6 \\
7500 & 15000 & 10 & 0.013 & 1.41e-5 & 4 & 38.3 &1.67e-4 & 68.0 \\
  \end{tabular}
\end{center}
\end{minipage}
\caption{\small The figure compares five matrix completion algorithms against GROUSE. The table gives a comparison of GROUSE and NNLS on large random matrix completion problems.}
\label{fig:mc-comparison}
\end{figure*}

In Figure~\ref{fig:mc-comparison} we show a comparison of five matrix completion algorithms and GROUSE.  Namely, we compare to the performance of  OPT-SPACE~\cite{Keshavan10a}, FPCA~\cite{Ma08}, SVT~\cite{Cai08}, SDPLR~\cite{Burer05,Recht07}, and NNLS~\cite{Toh09}. We downloaded each of these MATLAB codes from the original developer's websites when possible. We use the same random matrix model as in Figure~\ref{fig:mc-performance}.  GROUSE is faster than all other algorithms, and achieves higher quality reconstructions on many instances.  We subsequently compared against NNLS, the fastest batch method, on very large problems. Both GROUSE and NNLS achieved excellent reconstruction, but GROUSE was twice as fast.

\section{Discussion and Future Work}

The inherent simplicity and empirical power of the GROUSE algorithm merit further theoretical investigations to determine exactly when it provides consistent estimates.  It is possible that an adaptation of the analysis of~\cite{Keshavan10a} to this online setting will yield such consistency bounds.  The main difficulty lies in characterizing the trajectory of the GROUSE algorithm from a random starting subspace. Though we can guarantee that there is a basin of attraction around the global minimum, it is not yet clear how to characterize when GROUSE will end up in this basin.

We would also like to investigate how to adapt step-size in the GROUSE algorithm automatically to varying data.  This is the only parameter required to run the GROUSE algorithm, and we have seen that there are substantial performance gains when this parameter is optimized.  Using techniques from Least-Squares Estimation, it may be possible to automatically tune the step size based on our current error residuals.

\section{Acknowledgments}
This work was partially supported by the AFOSR grant FA9550-09-1-0140.  R. Nowak would also like to thank Trinity College and the Isaac Newton Institute at the University of Cambridge for support while this work was being completed.

\begin{small}
\bibliographystyle{abbrv}
\bibliography{GROUSE_Nov}
\end{small}

\end{document}